\documentclass[aps,graphicx,preprint, sort&compress]{revtex4-1}

\usepackage{graphicx}
\usepackage[percent]{overpic}
\usepackage{amssymb}
\usepackage{amsmath}
\usepackage{rotating}
\usepackage{colordvi}
\usepackage{bm}

\begin{document}

\title{Plasma wakefield excitation by incoherent laser pulses: 
a path towards high-average power laser-plasma accelerators}

\author{C.~Benedetti}
\affiliation{Lawrence Berkeley National Laboratory, Berkeley, California 94720, USA}
\author{C.~B.~Schroeder}
\affiliation{Lawrence Berkeley National Laboratory, Berkeley, California 94720, USA}
\author{E.~Esarey}
\affiliation{Lawrence Berkeley National Laboratory, Berkeley, California 94720, USA}
\author{W.~P.~Leemans}
\affiliation{Lawrence Berkeley National Laboratory, Berkeley, California 94720, USA}

\date[]{Submitted to Nature Publishing Group, 2013}

\maketitle

\section*{}
%%%%%%% INTRODUCTION
{\bf In a laser plasma accelerator (LPA), a short and intense laser
pulse propagating in a plasma drives a wakefield (a plasma wave with a
relativistic phase velocity) that can sustain extremely large
electric fields, enabling compact accelerating
structures \cite{rmp-esarey, natPhot-hooker, nature-leemans,
nature-downer}.  Potential LPA applications include compact radiation
sources \cite{prl-leemans-radiation, fuchs, maier_PRX, huang_PRL} and
high energy linear colliders \cite{physicstoday-leemans,
prstab-schroeder, prstab-nakajima}.   We propose and study plasma wave
excitation by an incoherent combination of a large number of low
energy laser pulses ({\sl i.e.}, without constraining the pulse
phases).  We show that, in spite of the incoherent nature of
electromagnetic fields within the volume occupied by the pulses,
the excited wakefield is regular and its
amplitude is comparable or equal to that obtained using a single,
coherent pulse with the same energy.  These results provide a path
to the next generation of LPA-based applications, where incoherently
combined multiple pulses may enable high repetition rate, high 
average power LPAs.}
%%%%%%%%%%%%

The plasma wave excited by a laser pulse propagating in an underdense
plasma is the result of the gradient in laser field energy density
providing a force (i.e., the ponderomotive force) that creates a space
charge separation between the plasma electrons and the neutralizing
ions.  For a resonant laser pulse, i.e., with a duration $T_0\sim
\lambda_p/c$, where $\lambda_p=2\pi/k_p=2\pi c/\omega_p$ is the plasma
wavelength, and $\omega_p=(4\pi n_0 e^2/m)^{1/2}$ is the electron
plasma frequency for a plasma with density $n_0$ ($c$ is the speed of
light in vacuum, $m$ and $e$ are, respectively, the electron mass and
charge), with a relativistic intensity, i.e., with a peak normalized
vector potential $a_0=eA_0/mc^2 \sim 1$ ($A_0$ is the peak amplitude
of the laser vector potential), the amplitude of the accelerating
field is of order $E_0=mc\,\omega_p/e$, or $E_0[\hbox{V/m}]\simeq
96\sqrt{n_0[\hbox{cm}^{-3}]}$.  For example, 
$n_0\sim 10^{17}$ cm$^{-3}$ gives
$E_{0}\simeq 30$ GV/m, which is three orders of magnitude
higher than in conventional accelerators.
%%%%%%%%%%%%

Applications for LPAs that produce electron beams with energies 1-10
GeV include free electron lasers in the x-ray regime
\cite{prl-leemans-radiation, fuchs, maier_PRX, huang_PRL} and
accelerator modules for linear colliders \cite{physicstoday-leemans,
prstab-schroeder, prstab-nakajima}.
For example, a conceptual design for a 1 TeV center-of-mass electron-positron
LPA-based linear collider is presented in
Ref.~\onlinecite{prstab-schroeder}.  
Both the electron and positron arms of the collider require 50 LPA stages  
at 10 GeV with a length of
$\sim 1$ m and a density $n_0\sim 10^{17}$ cm$^{-3}$.  Each
LPA stage is powered by a laser pulse with tens of J of energy
with duration $T_0\lesssim$ 100
fs
%wavelength $\lambda_0 \sim 1$ $\mu$m, 
%containing tens of Joules of
%size of the order of the plasma wavelength, $\sim\lambda_p$, yielding
and an intensity such that $a_0 \sim 1$, which is achievable with 
present laser technology.
% Every LPA stage provides an
% energy gain of $\sim 10$ GeV for particles properly phased in the
% plasma wave.  The final beam energy, after 50 stages, would then be
% 0.5 TeV. The total length of the accelerator, considering also the
% distance to couple-in the new laser pulses between stages, would be
% less than a km, to be compared with the $\sim 30$ km required using
% conventional accelerator technology\cite{website_ILC_CLIC}.  
The
required laser repetition rate, however, as dictated by luminosity requirements,
is $f_{rep} \sim 10$ kHz (an average power of hundreds of kW), which 
is orders of magnitude beyond present technology. 
%%%

To date, LPAs are typically driven by solid-state (e.g., Ti:sapphire) 
lasers that are limited to an average power $\sim 100$ W. For 
example, the Berkeley Lab Laser Accelerator (BELLA) laser delivers 
40 J pulses on target at 1 Hz \cite{LeemansPAC}. Since virtually all applications of LPA 
will benefit greatly from higher repetition rates, it is essential 
that high average power laser technology continues to be developed.

% operated with a repetition rate of $\lesssim 1$ Hz, approximately four
% orders of magnitude below the required $f_{rep}$.  Increasing the
% laser repetition rate from a few Hz to a few kHz will require
% development of new laser technologies able to deliver pulses with high
% peak power ($\sim$ PW) and, at the same time, operate at substantially
% higher average power ($\sim$ MW)%. 
% and with significantly higher wall-plug efficiency (10's of percents).
% We recall that the efficiency of current state-of-the-art petawatt
% Ti:Sapphire laser systems such as BELLA [1 Hz repetition rate, 40 J
% energy, 30 fs pulse duration] at the Lawrence Berkeley National
% Laboratory is $\ll 1$\%.
%%%%%%%%%%%%

Several laser technologies are being developed to provide systems 
with high average power and high efficiency \cite{natPhot-hooker, NatPhot_Mourou}.
% namely, fiber lasers, diode-pumped solid-state
% lasers, and optical parametric chirped pulse amplification based
% lasers.  
For example, in
Refs.~\onlinecite{NatPhot_Mourou, Proceeding_MourouHulinGalvanauskas},
a scheme is presented were a large number of diode-pumped fiber
systems, delivering pulses with $\sim$ mJ energy at kHz repetition
rate, are combined in such a way the relative phases of the output
beams are controlled so they constructively interfere (coherent
combination) and produce a single, high power output beam with high
efficiency.  Coherent combination of a large number of fiber lasers 
(e.g., $\sim 10^4$ fibers for a total pulse energy
of 10's of J) for an LPA is extremely challenging, requiring short pulse beams
 (duration $<$ ps) that are matched in phase, time and space.
% also the group velocity delays of different pulses using phase
% modulators and delay lines.  
To date, the coherent combination of an
array of 64 (continuous wave) beams from fiber amplifiers has been
demonstrated \cite{ICAN_64fibers}.  The coherent combination of a small
number of femtoseconds pulses has also been
achieved \cite{ICAN_64fibers_pulse1, ICAN_64fibers_pulse2}.
%%%%%%%%%%%%

As is shown below, an LPA does not require a fully coherent drive
laser pulse.  This is true because the LPA wakefield is excited by the
ponderomotive force (i.e., the gradient in the electromagnetic energy
density), along with the fact that the plasma responds on the time
scale $\lambda_{p}/c$.  Large amplitude wakefield excitation requires
sufficient electromagnetic energy within a given volume, typically of
the order of $\sim \lambda_p^3$.  Since the wakefield response behind
the driver depends on the time-integrated behavior of the
electromagnetic energy density of the driver over $\lambda_{p}/c$, it
is insensitive to time structure in the driver on time scales
$\ll\lambda_{p}/c$, which allows for the use of incoherently combined
laser pulses as the driver.  Theoretically, this can be easily
demonstrated in the linear ($a^{2}< 1$) wakefield regime in which
the amplitude of the electric field of the wake is small ($\vert {\bf
E}\vert/E_0< 1$).  In the linear regime, the wake electric field
${\bf E}$ is given by \cite{rmp-esarey}
\begin{equation}
\label{eq:eq_1}
\left({\partial^2}/{\partial t^2}+\omega_p^2\right ) {\bf 
E}/{E_0}=-({c \omega_p}/{2} )
\nabla a^{2}, 
\end{equation}
 with the solution 
\begin{equation}
{ {\bf E}}/{E_0}=-({c}/{2})\int_0^t dt' 
\sin[\omega_p(t-t')] \nabla a^2(t'),
\end{equation}
This Green function solution averages out the small 
scale
time structure in the ponderomotive force.
Hence, in effect, the wakefield is given by 
\begin{equation}
\label{eq:eq_3}
\left({\partial^2}/{\partial t^2}+\omega_p^2\right) 
{\bf  E}/{E_0}\simeq-({c \omega_p}/{2} )
\nabla  \langle a^{2}\rangle, 
\end{equation}
where the angular brackets represent a time average over scales small 
compared to $\lambda_{p}/c$.

% For instance, in the
% linear regime ({\sl i.e.}, for small wake amplitudes, namely
% $E_z/E_0\ll 1$), the longitudinal wakefield generated by a laser pulse
% described by the normalized vector potential $a(\rbf,t)$,
% is\cite{rmp-esarey}
% \begin{equation}
% \label{eq:eq_1}
% \left( \frac{\partial^2}{\partial t^2}+\omega_p^2\right ) \frac{E_z}{E_0}=-\frac{c \omega_p}{2} \frac{\partial a^2}{\partial z}.
% \end{equation}
%  The Green function solution to Eq.~(\ref{eq:eq_1}) reads 
% \begin{equation}
% \frac{ E_z(\rbf, t)}{E_0}=-\frac{c}{2}\int_0^t dt' \sin[\omega_p(t-t')] \frac{\partial a^2(\rbf, t')}{\partial z},
% \end{equation}
% and so the amplitude of the wakefield behind the driver is given by
% the convolution in time of the laser field with the Green function.

Owing to the time average process characterizing the wake excitation,
we show that multiple, low-energy,
incoherently combined laser pulses can deposit sufficient field energy in the
plasma to ponderomotively drive a large wakefield.  
We show that no phase
control in the combination of multiple laser pulses is required for
LPAs.  
We find that, under certain conditions, the wake
generated by an incoherent combination of pulses is regular behind the
driver and its amplitude is comparable, or equal, to the one obtained
by using a single coherent pulse with the same energy.  We expect that
the fundamental requirements to achieve incoherent combination are
more relaxed compared to coherent combination.
Hence, incoherent combination may provide an alternative and technically
simpler path to the realization of high repetition rate and high 
average power LPAs.
%%%%%%%%%%%%

\begin{figure}
\begin{overpic}[width=150mm,  tics=10]{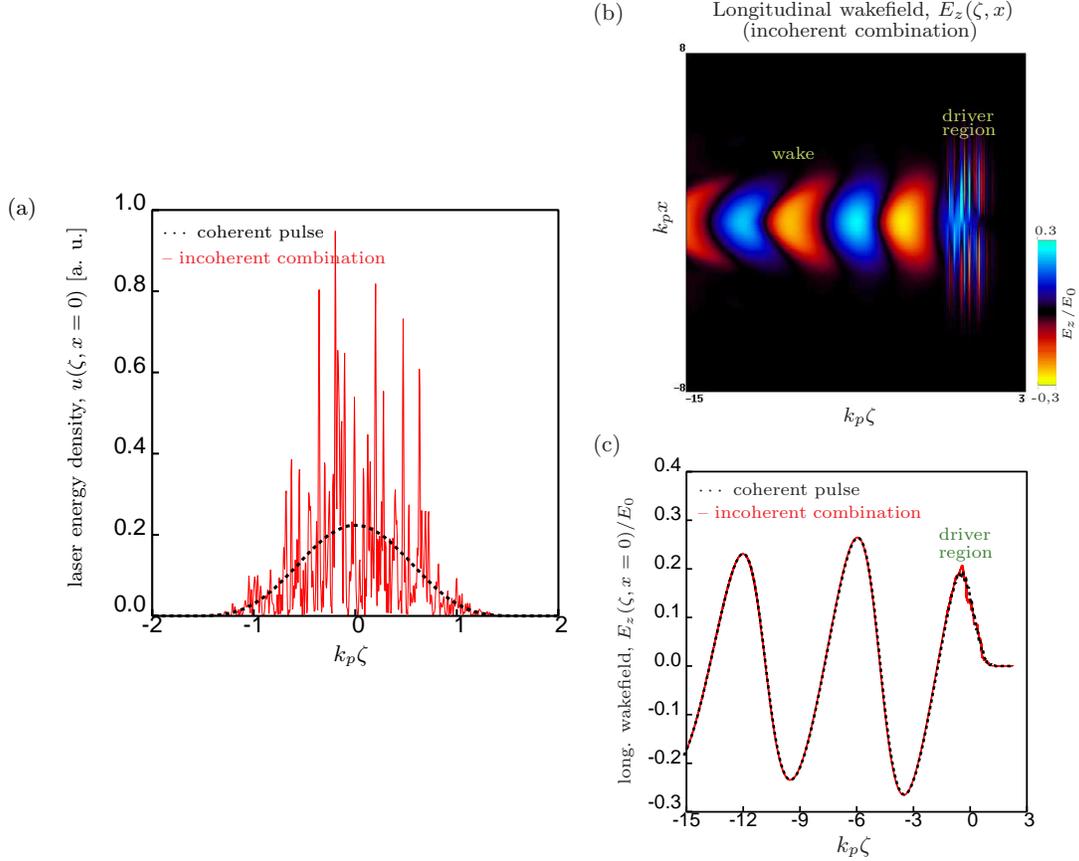}
\end{overpic}
%\begin{tabular}{cc}
%
%\begin{overpic}[width=90mm,  tics=10]{FIG1/fig_1a.ps}
%\put(4, 72) {(a)}
%\put (54.5,3) {$\footnotesize k_p \zeta$}
%\put(12,9) { \begin{sideways}\parbox{60mm} {\footnotesize  laser energy density, $u(\zeta, x=0)$ [a. u.]}\end{sideways} } 
%
%\put (28,68.5) {\scriptsize $\cdots$ coherent pulse}
%\put (28,64.5) {\Red{\scriptsize -- incoherent combination}}
%
%\end{overpic}
%
%&
%\raise 3.2 truecm
%\hbox{
%\begin{tabular}{c}
%\hskip -0.5 truecm
%\begin{overpic}[width=58.5mm,  tics=10]{FIG1/fig_1b.ps}
%\put (47,2.5) {\footnotesize$k_p\zeta$}
%\put(0,-2) { \begin{sideways}\parbox{62mm} {\footnotesize $k_px$ }\end{sideways} } 
%\put (8.7,95) {\footnotesize  Longitudinal wakefield, $E_z(\zeta, x)$}
%
%\put (70.5,74) {\scriptsize \SpringGreen{driver}}
%\put (70.5,70.8) {\scriptsize \SpringGreen{region}}
%
%\put(93,7) {\includegraphics[width=3mm, height=25mm]{colorbar_pos_neg.ps}}
%\put(91.8, 7.2) {\tiny -0,3} 
%\put(92.5,47) {\tiny 0.3} 
%\put(99,8.5) {\begin{sideways}\parbox{23mm} {\tiny $E_z/E_0$}\end{sideways} } 
%
%
%\end{overpic}
%
%\\
%\hskip -1.5 truecm
%\begin{overpic}[width=77mm,  tics=10]{FIG1/fig_1c.ps}
%\put(9, 153) {(b)}
%\put(9, 75) {(c)}
%\put (53,2.5) {$\footnotesize k_p \zeta$}
%\put(12,7) { \begin{sideways}\parbox{52mm} {\scriptsize long. wakefield, $E_z(\zeta, x=0)/E_0$}\end{sideways} } 
%
%\put (71.5,59) {\scriptsize \OliveGreen{driver}}
%\put (71.5,55.8) {\scriptsize \OliveGreen{region}}
%
%\put (28,67.) {\scriptsize $\cdots$ coherent pulse}
%\put (28,63.) {\Red{\scriptsize -- incoherent combination}}
%
%\end{overpic}
%
%\end{tabular}
%}
%
%\end{tabular}
%%%%%%%%%%%%%

\caption{{\bf Laser energy density and wakefield generated by a wavelength combining scheme.}
{\bf a}, On-axis lineout (red line) of the electromagnetic energy
density obtained overlapping 48 spectrally separated laser pulses
(including polarization multiplexing).  The field of each pulses has a
cosine-squared longitudinal profile with a FWHM duration of 30 fs, and
a Gaussian transverse profile with a spot size of 12 $\mu$m.  All the
pulses have the same energy.  The minimum and maximum laser
wavelengths of the pulses are $\lambda_{0, min}=0.2$ $\mu$m
($a_0(\lambda_{0,min})=0.03$), and $\lambda_{0, max}=2.7$ $\mu$m
($a_0(\lambda_{0,max})=0.47$).  The laser beams propagate (matched
propagation) in a parabolic plasma channel with an on-axis density
$n_0=10^{18}$ cm$^{-3}$.  The black dashed line is the electromagnetic
energy density for a single, circularly polarized, coherent, laser
pulse with an energy equal to that of the incoherently combined pulses,
and an intensity such that $a_{0, \hbox{\tiny coherent}}=1$ for
$\lambda_0=0.8$ $\mu$m.  {\bf b}, Map of the longitudinal wakefield,
$E_z(\zeta, x)$, generated by the incoherent combination. % after a propagation distance $s=0.1$ mm in the plasma.  
{\bf c}, On-axis lineout of the longitudinal wakefield for the incoherent
combination (red line) and for the coherent pulse (black dashed line).
\label{fig:fig_1}}
\end{figure}

To illustrate the physics of wake generation by incoherently combining
multiple laser pulses, we present two examples of incoherent
combination schemes, namely spectral combination and mosaic of
beamlets.
%%%%%%%%%%%%
In the first example we consider $N$ laser pulses with the same length
and spot size, and different (random) phases, propagating (along $z$
direction) in a parabolic plasma channel.  The laser spot size is the
one ensuring matched propagation in the channel.  The pulses have
different wavelengths and they are spectrally separated (i.e., the
power spectra of the pulses do not overlap with each other) and so the
total energy of the incoherent combination equals the sum of the
energies of the pulses.  From an experimental point of view, the
condition of spectral separation allows the overlap of the different
pulses by using a dispersive optical system like a sequence of
dichroic mirrors, a grating, or a prism \cite{review_laser_comb}.  In
Fig.~\ref{fig:fig_1}a we show (red line) the on-axis lineout of the
driver energy density generated by spectrally combining 48 pulses.
The pulses have the same energy (and intensity), and their centroids
are overlapped.  For each frequency in the combination we accommodate
two laser pulses with orthogonal polarizations (polarization
multiplexing).  The black dashed line in the same panel is the laser
energy density for a single (circularly polarized) coherent pulse with
the same energy (see Fig.~\ref{fig:fig_1} caption for the laser and
plasma parameters).  In Fig.~\ref{fig:fig_1}b we show a map, obtained
with the PIC code {\tt ALaDyn} \cite{aladyn1, aladyn2}, of the
longitudinal wakefield, $E_z(\zeta, x)$, $\zeta = z-ct$ being the
longitudinal coordinate relative to the laser driver and $x$ the
transverse coordinate, generated by the incoherent combination.  %after a propagation distance $s=0.1$ mm.
Finally, Fig.~\ref{fig:fig_1}c shows the on-axis lineout of $E_z$ for
the incoherent combination (red line) and for the circularly polarized
laser pulse (black dashed line).  We notice, from both
Fig.~\ref{fig:fig_1}b and Fig.~\ref{fig:fig_1}c, that the wakefield
from incoherent combination shows an incoherent pattern within the
driver region, namely for $|k_p\zeta|\lesssim 2$.  However, behind the
driver region ($k_p\zeta\lesssim -2$), the wakefield is regular and
its amplitude equals the one of the single coherent pulse with the
same energy.
%%%%%%%%%%%% %Laser and plasma parameters for this example are described in detail in the caption of Fig.~\ref{fig:fig_1}.  
\begin{figure}
%%%%%%%%%
\begin{overpic}[width=130mm,  tics=10]{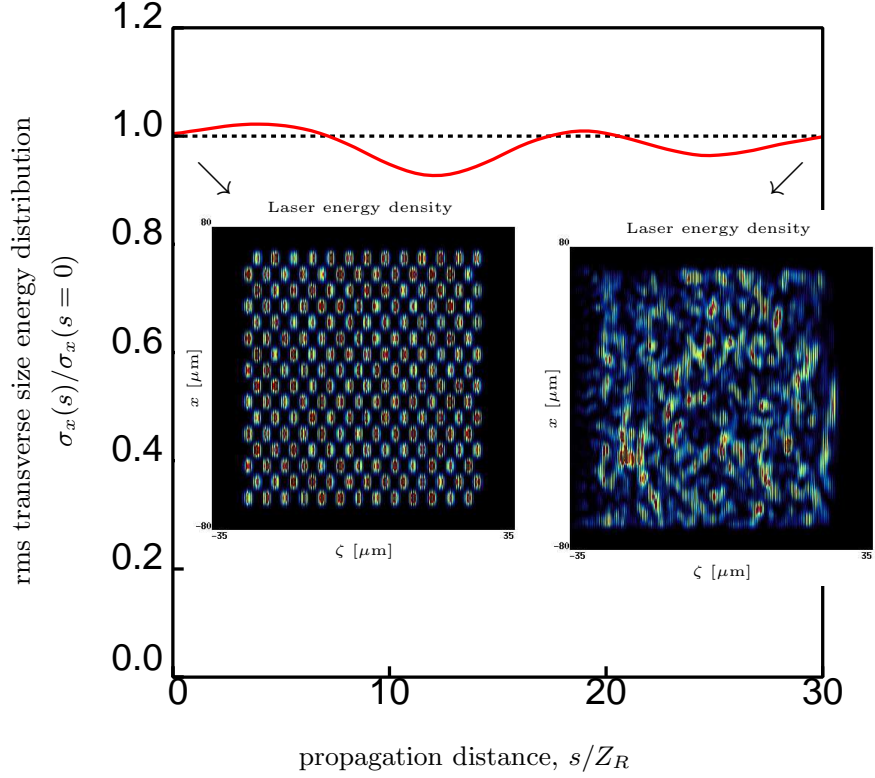}
\end{overpic}
%\begin{overpic}[width=120mm,  tics=10]{FIG2/fig_2.ps}
%\put (39,1.5) {propagation distance, $ s/Z_R$}
%\put(10,12) { \begin{sideways}\parbox{72mm} {rms transverse size energy distribution}\end{sideways} } 
%\put(14,12) { \begin{sideways}\parbox{72mm} {$\sigma_x(s)/\sigma_x(s=0)$}\end{sideways} } 
%
%\put(27, 21) {\includegraphics[width=44mm]{FIG2/fig_2insA.ps}}
%\put(29, 58){\large $\searrow$}
%\put(36, 55.5) {\tiny Laser energy density}
%\put(42.5, 22) {\tiny $\zeta$ [$\mu$m]}
%\put(27.2,31) { \begin{sideways}\parbox{20mm} {\tiny $x$ [$\mu$m]}\end{sideways} } 
%
%
%\put(62, 19) {\includegraphics[width=44mm]{FIG2/fig_2insB.ps}}
%\put(78, 56){\large $\swarrow$}
%\put(71, 53.5) {\tiny Laser energy density}
%\put(77.5, 20) {\tiny $\zeta$ [$\mu$m]}
%\put(62.2,29) { \begin{sideways}\parbox{20mm} {\tiny $x$ [$\mu$m]}\end{sideways} } 
%
%\end{overpic}
%%%%%%%%%%%%

\caption{{\bf Guiding of a mosaic of beamlets.}
Evolution of the rms transverse size of the energy distribution,
$\sigma_x(s)$, for a combinations of $208=13\times16$ identical
beamlets (the field of each beamlet has a cosine-squared dependence but
longitudinally and transversally) with $a_0=1.5$, $\ell_0=4$ $\mu$m
(total pulse length), $d_0=15$ $\mu$m (total pulse width),
$\lambda_0=0.8$ $\mu$m.  The background plasma density is
$n_0=0.9\cdot 10^{17}$ cm$^{-3}$.  The beamlets are tiling a 2D domain
$55$ $\mu$m $\simeq \lambda_p/2$ long and $144$ $\mu$m wide.
The two insets show time snapshots of the laser energy density at the
beginning of the simulation (left), and after some propagation
distance in the plasma (right), where the laser field exhibits an
incoherent pattern.
\label{fig:fig_2}}
\end{figure}

In the second example we consider a collection of short and narrow
laser beamlets placed side-by-side, both longitudinally and
transversally, tiling a prescribed volume.  Each pulse has high
(relativistic) peak intensity but low energy owing to the limited
spatial extent of the beamlets.  The beamlets, whose centroids are
arranged into an uniform spatial grid, have the same length
($\ell_0$), width ($d_0$), wavelength ($\lambda_0$) and intensity
($a_0$), and different (random) phases.  For each centroid location we
can accommodate two beamlets with orthogonal polarizations
(polarization multiplexing).
%The laser field for each beamlet is non-zero only over a finite domain, 
%determined by $\ell_0$ and $d_0$, around the pulse centroid
%
The laser field of each beamlet vanishes at a distance from the
centroid larger than $\ell_0/2$, longitudinally, and $d_0/2$,
transversally, and the fields of beamlets with the same polarization
do not overlap.  As a consequence, energy is additive (i.e., the total
energy is the sum of the energy of the single beamlets).  For
simplicity, in this example we will restrict the discussion to
two-dimensional (2D) Cartesian geometry.  The generalization to the
three-dimensional case is straightforward.  The guiding of the
incoherent combination of pulses over distances much longer compared
to the Rayleigh length of the single laser beamlets, namely $Z_R \sim
\pi d_0^2/\lambda_0$, can be achieved by a plasma channel with,
transversally, a constant density over the domain occupied by the
beamlets,
%up to a transverse distance equal to the extent of the beamlets, 
followed by steep plasma walls able to reflect and contain radiation
from diffracting beamlets.  Because of multiple reflections, and
interference between the fields of different beamlets, we expect the
total electromagnetic radiation driving the wake to have a complex
pattern.  In Fig.~\ref{fig:fig_2} we show the evolution of the rms
transverse size of the energy distribution, $\sigma_x(s)$, $s=ct$
being the propagation distance, for a combination of 208 beamlets
initial arranged in an array of $13\times16$ (see Fig.~\ref{fig:fig_2}
caption for the laser and plasma parameters).  We see that the
laser energy from the combination is well guided over distances
significantly longer than the Rayleigh length of the beamlets.  The
two insets show snapshots of the laser energy density at the beginning
of the simulation (initial beamlets configuration), and after some
propagation distance in the plasma, where the laser field exhibits a
clear incoherent pattern.
%%%%%%%%%%%%%%%%
\begin{figure}[!h]
\begin{overpic}[width=90mm,  tics=10]{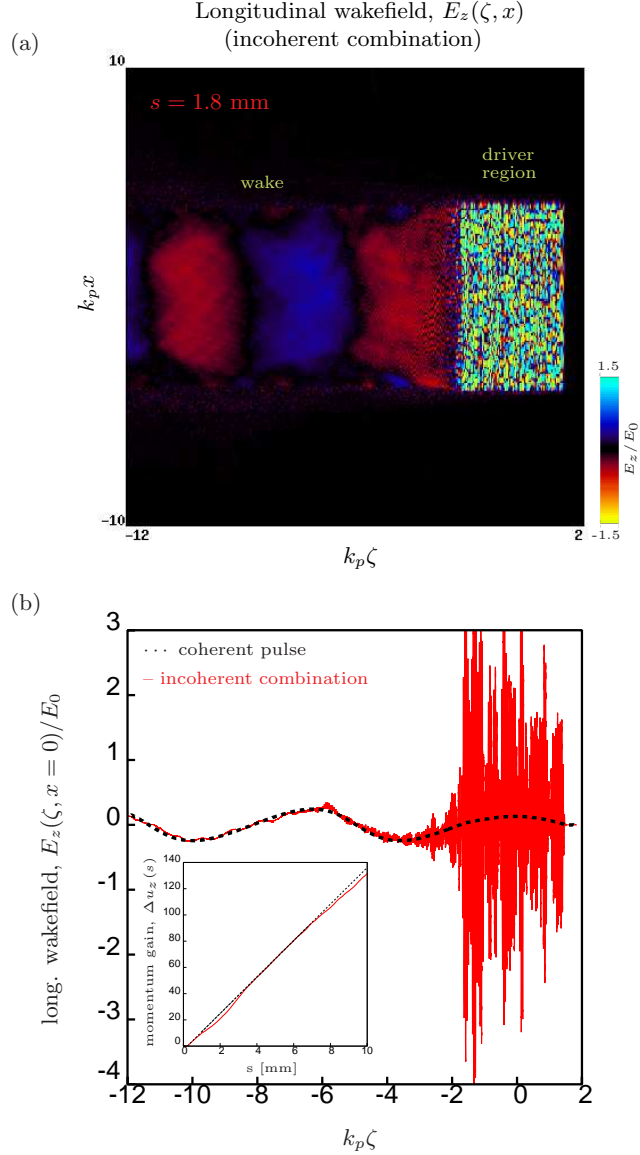}
\end{overpic}
%%%%%%%%%
%\begin{tabular}{c}
%\hskip 1.5 truecm
%\begin{overpic}[width=78.5mm,  tics=10]{FIG3/fig_3a.ps}
%\put (48,3) {$k_p\zeta$}
%\put(0,11) { \begin{sideways}\parbox{62mm} {$k_px$}\end{sideways} } 
%\put (20,95) {\footnotesize  Longitudinal wakefield, $E_z(\zeta, x)$}
%
%\put (14,83) {\WildStrawberry{$s=1.8$ mm}}
%
%\put(93,7) {\includegraphics[width=3mm, height=25mm]{colorbar_pos_neg.ps}}
%\put(91.8, 7.2) {\tiny -1.5} 
%\put(92.5,37) {\tiny 1.5} 
%\put(97.5,8.5) {\begin{sideways}\parbox{23mm} {\tiny $E_z/E_0$}\end{sideways} } 
%
%\end{overpic}
%
%\\
%\hskip .15 truecm
%\begin{overpic}[width=102mm,  tics=10]{FIG3/fig_3b.ps}
%\put(10, 150) {(a)}
%\put(10, 74) {(b)}
%\put (55,1.5) {$ k_p \zeta$}
%\put(13,10) { \begin{sideways}\parbox{62mm} { long. wakefield, $E_z(\zeta, x=0)/E_0$}\end{sideways} } 
%
%\put (28,68) {\scriptsize $\cdots$ coherent pulse}
%\put (28,64) {\Red{\scriptsize -- incoherent combination}}
%%\put (28,61) {\Red{\scriptsize $\phantom{x} $(208 beamlets)}}
%
%
%\put(23, 11) {\includegraphics[width=40mm]{FIG3/fig_3bins.ps}}
%\put(27, 13) { \begin{sideways}\parbox{30mm} {\tiny momentum gain, $\Delta u_z (s) $}\end{sideways} } 
%\put(42, 11.5) {\tiny s [mm]}
%
%\end{overpic}
%
%\end{tabular}
%%%%%%%%%%%%

\caption{{\bf Wakefield generated by a mosaic of beamlets.}
{\bf a}, Map of the longitudinal wakefield, $E_z(\zeta, x)$, generated
by the incoherent combination of 208 beamlets (same parameters as in
Fig.~\ref{fig:fig_2}) after a propagation distance $s=1.8$ mm in the
plasma.  {\bf b}, On-axis lineout of the longitudinal wakefield for
the incoherent combination (red line) and for a single coherent pulse
with $a_{0, \hbox{\tiny coherent}}=0.8$ and same length and rms spot
size as the incoherent combination (black dashed line).  This inset
shows the integrated momentum gain, $\Delta u_z(\zeta, s) \simeq -
(e/mc^2) \int_0^s E_z(\zeta, x=0, s')ds'$, for a relativistic particle
initially located in $k_p\zeta\simeq -10$ (maximum accelerating
field).  The black and red lines refer, respectively, to the momentum
gain in the coherent and incoherent case.  \label{fig:fig_3}}
\end{figure}

The wakefield generated by the mosaic of incoherent beamlets is
presented in Fig.~\ref{fig:fig_3}.  The laser and plasma parameters
are the same as in Fig.~\ref{fig:fig_2}.  In Fig.~\ref{fig:fig_3}a we
show a 2D map of the longitudinal wakefield generated by the
incoherent combination after a propagation distance $s=1.8$ mm.  In Fig.~\ref{fig:fig_3}b we show the on-axis
lineout of the longitudinal wakefield for the incoherent combination
(red line) and for a single coherent pulse (uniform transverse intensity profile) with an intensity such that $a_{0, \hbox{\tiny
coherent}}=0.8$, and with the same length and rms spot size as the
incoherent combination (black dashed line).  We notice that, behind
the driver region, the wake from incoherent combination is regular and
its amplitude is the same as the one from a single coherent pulse.
The noisy field structure observed in the lineout of $E_z$, due to
multiple reflections from walls and interference of beamlets, does not
affect the energy gain of relativistic particles accelerated in the
wakefield.  This is shown in the inset of Fig.~\ref{fig:fig_3}b, where
we compute the integrated momentum gain as a function of the
propagation distance for a relativistic particle initially located in
$k_p\zeta\simeq -10$ (maximum accelerating field).  The black and red
lines in the inset refer, respectively, to the momentum gain in the
coherent and incoherent case.  No significant difference is observed
in the momentum gain in the two cases ($\sim 2 \%$ difference in the energy gain after 10 mm propagation).  The total energy of the
combination of pulses exceeds the one of the coherent pulse by $\sim 10 \%$.  This (moderate) loss
in the efficiency of the combination can be partially ascribed to the
fact that the varying transverse intensity profile within each beamlet
results in an effective less-than-unity fill factor.  
%Note also that here we are comparing a collection of pulses forming a transversally flat-top intensity profile with a single pulse with a Gaussian intensity profile.

%%%%%%%%%%%%

In summary, we have shown that it is possible to efficiently excite a
wakefield for LPA applications with an incoherent combination of
multiple, low-energy, laser pulses.  Since no phase control is
required, we expect that the fundamental requirements to achieve
incoherent combination are more relaxed compared to coherent
combination, thereby enabling a technologically
simpler path for design of high-average power,
high-repetition rate LPA applications.

%\section*{References}

%%%%%%%%%%%%%%%%%%%%%%%%%%%%%%%%%%%%%%%%%%%%%%%%%%%%%%%%%%%%%%%%%%%%%%%%%%
%%%%%%%%%%%%%%%%%%%%%%%%%%%%%%%%%%%%%%%%%%%%%%%%%%%%%%%%%%%%%%%%%%%%%%%%%%

%\begin{acknowledgments}
\section*{Acknowledgments}
This work was first presented by C. Benedetti at the APS Division of Plasma Physics Meeting, Denver, CO, November 12, 2013, and subsequently submitted for publication.

This work was supported by the Director, Office of Science, Office of
High Energy Physics, of the U.S. DOE under Contract No.
DE-AC02-05CH11231, and used the computational facilities (Hopper,
Edison) at the National Energy Research Scientific Computing Center
(NERSC).  We would like to thank A. Galvanauskas for useful
discussions and suggestions.
%\end{acknowledgments}

%\section*{Author contributions}
%E.E. proposed the concept of using incoherent combination of laser pulses for wake excitation, which was further developed by C.B., C.B.S., E.E. and W.P.L. 
%C.B. performed the detailed analytical and numerical study of the problem, using a numerical code that he developed.


\begin{thebibliography}{99}

\bibitem{rmp-esarey} Esarey, E., Schroeder, C.~B. \& Leemans, W.~P. Physics of laser-driven plasma-based electron accelerators..  {\sl Rev. Mod. Phys.} {\bf 81,} 1229-1285 (2009).

\bibitem{natPhot-hooker} Hooker,  S. Developments in laser-driver plasma accelerators. {\sl Nature Photon.} {\bf 7,} 775-782 (2013).

\bibitem{nature-leemans} Leemans, W.~P. {\sl et al.} GeV electron beams from a centimetre-scale accelerator. {\sl Nature Phys.} {\bf 2,} 696-699 (2006).

\bibitem{nature-downer} Wang, X. {\sl et al.} Quasi-monoenergetic laser-plasma acceleration of electrons to 2GeV. {\sl Nat Commun.} {\bf 4,} 1988 (2013).

\bibitem{prl-leemans-radiation} Leemans, W.~P. {\sl et al.} Observation of terahertz emission from a laser-plasma accelerated electron bunch crossing a plasma-vacuum boundary. {\sl Phys. Rev. Lett.} {\bf 91,} 074802 (2003).

\bibitem{fuchs} Fuchs, M. {\sl et al.} Laser-driven soft-x-ray undulator source. {\sl Nature Phys.} {\bf 5,} 826-829 (2009).

\bibitem{maier_PRX} Maier, A.~R. {\sl et al.} Demonstration Scheme for a Laser-Plasma-Driven Free-Electron Laser. {\sl Phys. Rev. X} {\bf 2,} 031019 (2012).

\bibitem{huang_PRL} Huang, Z., Ding, Y., \& Schroeder, C.~B. Compact X-ray free-electron laser from a laser-plasma accelerator using a transverse-gradient undulator. {\sl Phys. Rev. Lett.} {\bf 109,}  204801 (2012).

\bibitem{physicstoday-leemans} Leemans, W.~P. \& Esarey, E. Laser-driven plasma-wave electron accelerators. {\sl Phys. Today} {\bf 62,} 44-49 (2009).

\bibitem{prstab-schroeder} Schroeder, C.~B., Esarey, E., Geddes, C.~G.~R., Benedetti, C. \& Leemans, W.~P. 
Physics considerations for laser-plasma linear colliders. {\sl Phys. 
Rev. ST Accel. Beams} {\bf 13,} 101301 (2010);
Schroeder, C.~B., Esarey, E.,  \& Leemans, W.~P. 
 {\sl Phys. Rev. ST Accel. Beams} {\bf 15,} 051301 (2010).

\bibitem{prstab-nakajima} Nakajima, K. {\sl et al.} Operating plasma density issues on large-scale laser-plasma accelerators toward high-energy frontier. {\sl Phys. Rev. ST Accel. Beams} {\bf 14,} 091301 (2011).

%\bibitem{website_ILC_CLIC} {\tt http://www.linearcollider.org/}

\bibitem{LeemansPAC} W.P. Leemans {\it et al.}, Particle Accelerator 
Conference Proc. (2013).

\bibitem{NatPhot_Mourou} Mourou, G., Brocklesby, B., Tajima, T. \& Limpert, J. The future is fibre accelerators. {\sl Nature Photon.} {\bf 7,} 258-261 (2011).

\bibitem{Proceeding_MourouHulinGalvanauskas} Mourou, G. A.,  Hulin, D. \& Galvanauskas, A. The road to high peak power and high average power lasers: Coherent-Amplification-Network (CAN). {\sl AIP Conf. Proc.} {\bf 827,} 152-163 (2006).

\bibitem{ICAN_64fibers} Bourderionnet, J., Bellanger, C., Primot, J.  \& Brignon, A. Collective coherent phase combining of 64 fibers. {\sl Opt. Express} {\bf 19,} 170530-17058 (2011).

\bibitem{ICAN_64fibers_pulse1} Daniault, L. {\sl et al.} Passive coherent beam combining of two femtosecond fiber chirped-pulse amplifiers. {\sl Opt. Lett.} {\bf 36,} 4023-4025 (2011).

\bibitem{ICAN_64fibers_pulse2} Klenke, A. {\sl et al.} Coherently-combined two channel femtosecond fiber CPA system producing 3 mJ pulse energy. {\sl Opt. Express} {\bf 19,} 24280-24285 (2011).

\bibitem{review_laser_comb} Fan, T.~Y. Laser beam combining for high-power, high-radiance sources. {\sl IEEE Journal of Selected topics in quantum electronics} {\bf 11,}  567-577 (2005).

\bibitem{aladyn1} Benedetti, C., Sgattoni, A., Turchetti, G. \& Londrillo, P. ALaDyn: A high-accuracy PIC code for the Maxwell-Vlasov equations. {\sl IEEE Transactions on plasma science} {\bf 36,} 1790-1798 (2008).

\bibitem{aladyn2}  Benedetti, C. {\sl et al.} PIC simulations of the production of high-quality electron beams 
via laser-plasma interaction. {\sl Nucl. Instr. and Meth. A} {\bf 608,} S94-S98 (2009).

\end{thebibliography}
\end{document}